%% file: main.tex
\documentclass[conference]{IEEEtran}
\usepackage{todonotes}
\usepackage{hyperref}
\usepackage{cleveref}
\usepackage{xcolor}
\usepackage{booktabs}

\definecolor{pos}{HTML}{000000} 
\definecolor{neg}{HTML}{000000} 

\ifCLASSINFOpdf
\else
\fi
\begin{document}
%
\title{Analyzing the Difficulty of Programming Assignments with Interpretable Knowledge Component Metrics}




%
\author{\IEEEauthorblockN{Tsvetomila Mihaylova\IEEEauthorrefmark{1},
Jing Fan\IEEEauthorrefmark{1},
Bita Akram\IEEEauthorrefmark{2}, 
Narges Norouzi\IEEEauthorrefmark{3} \\
Peter Brusilovsky\IEEEauthorrefmark{4},
Juho Leinonen\IEEEauthorrefmark{1} and
Arto Hellas\IEEEauthorrefmark{1}}
\IEEEauthorblockA{\IEEEauthorrefmark{1}Department of Computer Science, 
Aalto University,
Espoo, Finland, tsvetomila.mihaylova@aalto.fi}
\IEEEauthorblockA{\IEEEauthorrefmark{2}North Carolina State University, USA}
\IEEEauthorblockA{\IEEEauthorrefmark{3}University of California, Berkeley, USA}
\IEEEauthorblockA{\IEEEauthorrefmark{4}University of Pittsburgh, USA}}


\maketitle

\begin{abstract}
\input{00-abstract}
\end{abstract}


%
\IEEEpeerreviewmaketitle

\input{10-introduction}
\input{20-background}

\input{30-methodology}
\input{40-results}
\input{50-discussion}
\input{60-conclusion}

\section*{Acknowledgment}
The authors thank Professor Lisa Zhang of the University of Toronto Mississauga for providing course assignments used in this analysis.

This work was supported by Research Council of Finland grants \#356114 and \#367787.



%
\bibliographystyle{IEEEtran}
\bibliography{refs}

\end{document}

%% file: 00-abstract.tex
This full research paper examines how Knowledge Components (KCs) - fine-grained concepts or skills required to solve programming tasks - can be used as interpretable signals for understanding assignment difficulty and student struggle in introductory programming courses. While prior work has focused on predictive models based on programming behavior, such models are often difficult to interpret and therefore hard to use for instructional decisions. We take a complementary approach by analyzing KC-based metrics, including the number of KCs per assignment and changes in KC coverage between consecutive assignments (such as newly introduced or removed KCs). We study how these metrics relate to student performance, submission behavior, and engagement across a full course. The KCs can be given by the course instructor or extracted by using a large language model (LLM). We examine correlations between the number of KCs and student performance on the assignment, and analyze changes in KCs across assignments to identify cases where performance declines without new concepts being introduced. Selected assignments are then qualitatively inspected to understand potential design issues. Our results on data from three introductory programming course datasets show that assignments involving more KCs are generally associated with lower performance, and that sudden shifts in required KCs can coincide with disruptions in learning progression. We also identify assignments where performance declines even though no new KCs are introduced, suggesting potential issues in task design or instruction, which could be examined with qualitative analysis. In addition, we highlight which KCs are consistently associated with greater difficulty throughout the course. We propose an interpretable framework for analyzing programming assignments using KC-based metrics, with practical implications for instructors and course designers who want to better understand where and why students struggle, and how course materials might be improved. Our method can use KCs defined by an expert or extracted by an LLM, offering instructors an additional way to assess assignment quality beyond average correctness. It can be applied to any course with ordered assignments and measurable performance.


%% file: 10-introduction.tex
\section{Introduction}
\label{section:introduction}

Designing effective programming assignments is a key part of computing education. Yet, instructors often lack clear, interpretable indicators to assess whether an assignment is too difficult, conceptually misaligned, or poorly sequenced. While considerable research has focused on identifying struggling students~\cite{hellas2018predicting} through programming aptitude tests and broader psychological and background factors~\cite{evans1989best, shute1991likely}, behavioral modeling~\cite{ahadi2015exploring,carter2015normalized,jadud2006methods,watson2013predicting}, or performance in classroom activities~\cite{liao2016lightweight}, these approaches are typically student-centered and the potential interventions are less scalable in large courses. Moreover, while data-driven models can predict individual risk, they often lack transparency, limiting their use in diagnosing and improving instructional content.


An alternative and complementary approach is to examine the assignments themselves, specifically focusing on the cognitive demands they place on students. Prior work has explored metrics such as time-on-task~\cite{ihantola2014automatically} or code complexity~\cite{bastian2022comparing} to estimate assignment difficulty, but such indicators are not always informative for instructional design. In addition to understanding that an assignment is difficult, instructors need to understand why it is difficult.


We propose using Knowledge Components (KCs)--fine-grained concepts or skills required to complete specific tasks--as interpretable signals of assignment complexity. KCs offer a middle ground between high-level intended learning outcomes and concrete assignments, enabling a structured analysis of the conceptual content embedded in programming exercises. KCs have long been used in intelligent tutoring systems to model student knowledge~\cite{rivers2016learning,abdelrahman2023knowledge}, but their potential to inform assignment design remains underexplored in computing education.
KCs can be manually given by course designers, or can be extracted by processing Abstract Syntax Trees~\cite{hosseini2013JavaParser,demirtas2024reexamining}, by data-driven methods such as code embeddings and deep learning \cite{shi2023kc}, or by prompting large language models (LLMs) \cite{niousha2024use,fan2025adaptive}.

\begin{figure*}
    \centering
    \includegraphics[width=0.9\linewidth]{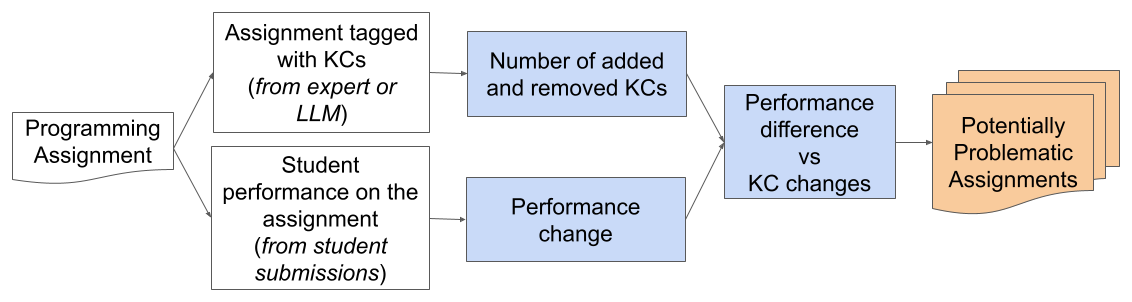}
    \caption{Proposed pipeline for assignment analysis using knowledge components.}
    \label{fig:pipeline}
\end{figure*}



In this work, we study whether KC-based metrics can serve as useful proxies for assignment difficulty and provide actionable insights for course designers. We focus on two key research questions:

\begin{description}
    \item[\textbf{RQ1:}] How does performance on programming assignments relate to the number of KCs in each exercise?
    \item[\textbf{RQ2:}] How can the number and change of KCs across assignments be used to evaluate the quality of learning materials?
\end{description}

While prior work has primarily used KCs for modeling student knowledge and prediction, we shift the focus to the assignment level, treating KCs as interpretable descriptors of conceptual load. This enables a complementary perspective where KCs are used not to model learners, but to diagnose and improve the structure and sequencing of instructional materials.

Our contributions are as follows:

\begin{itemize}
    \item We introduce a novel application of knowledge component analysis for evaluating assignment difficulty. 
    \item We empirically demonstrate that the number of KCs in an assignment correlates with student performance.
    \item We propose a new framework for detecting problematic assignments (\autoref{fig:pipeline}) by identifying performance drops that are not explained by the introduction of new KCs. \footnote{We have made a simple tool which practitioners can use to apply our method for their own courses: \url{https://kc-assignment-analyzer.lovable.app/}. The code can be found in \url{https://github.com/tsvm/kc-assignment-analyzer/}}
\end{itemize}

%% file: 20-background.tex
\section{Background}
\label{section:background}

Computing education researchers have long been interested in identifying struggling students~\cite{hellas2018predicting}. Early work focused on psychological and background factors as predictors of programming aptitude~\cite{evans1989best, shute1991likely}. More recent approaches have emphasized behavioral indicators such as time-on-task~\cite{leinonen2022time}, compilation behavior~\cite{jadud2006methods,watson2013predicting,carter2015normalized}, and performance in classroom activities such as peer instruction~\cite{liao2019robust}. Although these methods are useful for early prediction and intervention at the student-level, they fall short of guiding instructional design. For example, knowing that a student is likely to fail does not reveal whether certain assignments are misaligned or whether instructional scaffolding is insufficient.

A related but less extensively studied topic is estimating the difficulty of individual assignments. Understanding which exercises are likely to pose challenges is valuable for instructional design, enabling more effective scaffolding and pacing. Methods such as Item Response Theory~\cite{santos2020systematic} and Performance Factor Analysis~\cite{tirronen2020estimating} have been used to estimate difficulty in educational contexts, especially for multiple-choice questions. 
Other studies have explored task difficulty through programming-specific indicators such as code complexity metrics or expert judgement~\cite{bastian2022comparing}. Although scalable, these approaches tend to offer limited insight into the specific conceptual challenges students may face. Metrics such as time-on-task~\cite{ihantola2014automatically} can provide direct signals of difficulty, but may conflate conceptual understanding with student persistence or strategy. 

Previous work has analyzed programming exercises by extracting minimal programming constructs from student submissions to compute similarity and support exercise clustering or sequencing \cite{pan2025grouping}. In contrast, current work uses instructor- or LLM-defined KCs as interpretable proxies for conceptual load, focusing on diagnosing assignment difficulty and identifying structurally problematic exercises rather than measuring exercise similarity. 

Some assignments may be challenging due to the underlying conceptual load. The theory of threshold concepts--ideas that are transformative but difficult to grasp--has been influential in explaining why certain programming topics, such as recursion or abstraction, cause difficulties for students~\cite{meyer2003threshold,sanders2016threshold}. Recognizing the conceptual source of difficulty is crucial for assignment design, but it is difficult to infer from behavioral metrics alone.

Knowledge Components (KCs) are fine-grained concepts or skills that students are expected to master in a course~\cite{koedinger2012knowledge}. They are widely used in intelligent tutoring systems and in knowledge tracing models to track students' evolving knowledge~\cite{rivers2016learning,ai2019concept, abdelrahman2023knowledge}. In programming education, a KC might represent a specific construct (e.g. ``for-loop'') or a coarser action that would be composed of multiple specific KCs and their application (e.g. ``looping with a sentinel condition'').

Concept-level representations have also been explored to support instructional design decisions, for example, through design analytics tools that visualize concept coverage to help instructors balance learning activities during course planning~\cite{albo2019concept}.

Manual assignment of KCs for each assignment can be a tedious and time-consuming task for instructors. 
Therefore, efforts have been done to automate this process.
Earlier approaches to KC extraction in programming heavily relied on analyzing Abstract Syntax Trees (ASTs) to identify programming constructs and patterns~\cite{hosseini2013JavaParser,demirtas2024reexamining}. These approaches were widely adopted prior to the use of LLMs.
Recent works have demonstrated that it is possible to extract KCs from programming assignments using data-driven methods such as code embeddings and deep learning \cite{shi2023kc}, as well as more recent approaches based on LLMs \cite{shi2024knowledge, niousha2024use, fan2025adaptive}. It has been shown that the LLM-extracted KCs could have a similar quality to the ones manually assigned from an instructor \cite{fan2025adaptive}.

Perhaps because they are labor-intensive to create~\cite{aleven2013knowledge}, while KCs are widely used in adaptive learning and student modeling, they have been underutilized in evaluating the structure and sequencing of programming assignments. Most existing work treats assignments as atomic units, ignoring the granular concept load embedded within them, and focuses on outcomes (e.g. correctness or dropout) without considering why particular exercises may be more difficult.

In this work, we address this gap by exploring the relationship of KCs and assignment difficulty. Our objective is not to predict student performance per se, but to use KCs as an interpretable proxy for assignment difficulty and to identify potentially problematic exercises. Our approach is applicable to KCs assigned by an instructor or extracted by LLMs.

%% file: 30-methodology.tex
\section{Methodology}
\label{section:methodology}

\subsection{Problem Formulation}

For a programming course, we have the following information:

\begin{itemize}
    \item $\mathcal{A} = [a_1, ... , a_{c_a}]$ - Ordered list of \textit{assignments} in a course. The order of the assignments is the one in which they were presented in the course.
    \item $\mathcal{S} = [s_1, ... , s_{c_s}]$ - Ordered list of \textit{submissions} to these assignments for all students. The order is according to the time in which they were submitted in the system. 
    \item $\mathcal{K} = \{k_1, ... , k_{c_k}\}$ - Set of \textit{KC types}. For the Dart-Intro dataset, which is the main dataset used in this work, these are Fine-Grained KCs and Key Learnables.
    \item $k_i = \{kc_i^1, ... , kc_i^{c_{ki}}\}$ - Set of \textit{KCs} for each KC type.
    \item $\mathcal{U} = \{u_1, ... , u_{c_u}\}$ - Set of \textit{users} who make submissions to the assignments. We sometimes refer to the users as \textit{students}, but the terms are used interchangeably. 
\end{itemize}
In the above definitions, $c_a$, $c_s$, $c_k$ and $c_u$, indicate the count of the elements.

For each \textbf{submission} $s_j \in [s_1, ... , s_{c_s}]$, we receive the following:

\begin{itemize}
    \item The assignment $a_{s_j}$ for which it was submitted.
    \item The user $u_{s_j}$ who made the submission.
    \item Whether the submission is \textit{correct} or \textit{incorrect}.
\end{itemize}
Each student can make more than one submission to each assignment, and each separate submission is marked as either correct or incorrect. 




For an \textbf{assignment} $a_i \in [a_1, ... , a_{c_a}]$, we have the following:

\begin{itemize}
    \item Set of KCs for each KC type. We denote $NKC(a_i, k_j)$ as the number of KCs from type $k_j$ in assignment $a_i$.
    \item Assignment instructions, consisting of the task description and the starter code. There is also a solution code, which is not visible to the user.
    \item Number of submissions ($NS(a_i)$). We denote as $NS_{correct}(a_i)$ and $NS_{incorrect}(a_i)$ as the number of correct and incorrect submissions for assignment $a_i$.
    \item Number of users ($NU(a_i)$) who submitted this assignment.
    \item Assignment performance, which is measured as the average correctness ($AC(a_i)$), i.e., the fraction of the correct submissions for this assignment:

    \begin{equation}
        AC(a_i) = \frac{NS_{correct}(a_i)}{NS(a_i)}
    \end{equation}

    \item Difference of average correctness, number of submissions and number of users for the assignment $a_i$ is the change in these values since the previous assignment $a_{i-1}$. For the first assignment these values are zero. The difference in the average correctness is:
    
    \begin{equation}
        AC_{Diff}(a_i) = AC(a_i) - AC(a_{i-1})
    \end{equation}

    \item Change in KCs since the previous assignment. $NKC_{added}(a_i, k_j)$ and $NKC_{removed}(a_i, k_j)$ show how many KCs of KC type $k_j$ were added or removed from the KC set of type $k_j$ since the previous assignment $a_{i-1}$.
    
\end{itemize}

For each KC $kc_j$, we define:

\begin{itemize}
    \item $NS(kc_j)$ as the number of submissions for $kc_j$ for all assignments with which $kc_j$ is present. We denote $NS_{correct}(kc_j)$, $NS_{incorrect}(kc_j)$ as the number of correct and incorrect submissions for the KC $kc_j$. 
    \item $AC(kc_j)$ as a measure of KC performance using the average correctness of $kc_j$ defined as the fraction of correct submissions for $kc_j$:
    
    \begin{equation}
        AC(kc_j) = \frac{NS_{correct}(kc_j)}{NS(kc_j)}
    \end{equation}

\end{itemize}

\subsection{Data}


We use three datasets in our evaluation. We use one main dataset for detailed evaluation, and two additional datasets for verifying the generalizability of the approach.

\textit{Dart-Intro} is a dataset with assignments in introductory programming with the Dart programming language. The assignments in this dataset are from an online course by Aalto University, Finland.
The dataset consists of 45 assignments, split into 6 modules. Each assignment contains a handout, starter code, solution code, and KCs on two levels: fine-grained KCs (96 KCs) and key learnables (23 KCs). The KCs were manually assigned by the instructor of the course. 
The dataset includes 289,450 submissions to these assignments, made by 6,955 students. For each submission, we are given the assignment, user, order of submission for this assignment, submission code, and whether the submission was marked as correct or incorrect based on the automatic tests.
Therefore, for each assignment, we can calculate the average correctness, number of submissions, and number of users who tried the assignment. Each student can make more than one submission per assignment. 

\textit{Data-Structures} is a dataset from assignments used in a Data Structures course from University of Toronto Mississauga, Canada. It consists of lectures and labs, and has 56 assignments, formulated as multiple choice questions with one correct answer, and few questions with more than one correct answer. The assignments have a specific order and build on top of each other. For this course, we do not have KCs given by the instructor, but we extract them using an LLM.

To validate the assumption of correlation between the number of KCs and assignment performance, we use the ITAP Goal dataset\footnote{\url{https://pslcdatashop.web.cmu.edu/DatasetInfo?datasetId=1801}} \cite{rivers2016learning, rivers2017data}, which contains submissions for introductory programming in Python. It contains the following KC models: Intro CS Concepts (10 KCs), Tokens (48 KCs), Tokens Merged (35 KCs).
This is a public dataset, and the content of the assignments and their order in the course are not available. Therefore, we use this dataset only for validating the dependencies between the average correctness and number of KCs, but not for the experiments which require order of the assignments.

\subsubsection{KC Extraction with LLMs}

For the Dart-Intro and Data-Structures datasets, we extract KCs using GPT-5. We use a two-step prompting process. First, we prompt the LLM with all the assignments and ask for generation of list of KCs from all assignments. Then, we prompt again, passing the assignments together with the generated KC list, and prompt the model to tag each assignment with KCs from the list. The LLM sometimes asks confirmation questions, which were answered before obtaining the final lists. We use the final KCs in our experiments, the same way as we use the KCs given by instructors.

\subsection{Experiments}
We conduct a series of experiments to explore how KCs relate to student performance. Our goals are to (1) examine correlations between KC counts and performance metrics, (2) identify assignments with unexpected performance drops despite stable KC coverage, and (3) detect the most challenging KCs based on correctness trends.
\subsubsection{Experiment 1: Correlation between the Number of KCs and Performance on the Assignment.} With this experiment, we aim to answer RQ1. We calculate the Pearson correlation between the predictor and the specified target variables as described below. We split this experiment into two parts, exploring different dependencies:

\paragraph{Experiment 1.1.} \textit{How does the number of KCs affect the performance on the assignment?} For this experiment, we calculate the Pearson correlation between the number of KCs and the assignment performance $AC(a_i)$.
\paragraph{Experiment 1.2.} \textit{How does change of KCs change the performance on the assignment?} For this experiment, we calculate the correlations between the number of removed and added KCs and the change in the assignment performance $AC_{Diff}(a_i)$.


\subsubsection{Experiment 2: Identification of Potentially Problematic Assignments using Changes in KCs.}

This experiment addresses RQ2, by identifying assignments which do not introduce new KCs, but the performance on them drops from the previous assignment.
The hypothesis is that if the previous assignment was addressing certain of KCs, and the students managed to solve it, they have acquired some knowledge of those KCs and the performance on the subsequent assignments with the same KCs should be easier for them, i.e., we expect that the average correctness would not drop. 
Thus, \textit{there might be potential problems with the assignment for which there are no new KCs added, but the average correctness drops}.
After identifying such assignments, we manually review the student submissions on them to identify patterns of common errors that might indicate a problematic formulation or unexpected difficulty of the assignment. The pipeline for detecting the potentially problematic assignments is described in \autoref{fig:pipeline}.

As an addition to assignment analysis, we want to identify the most challenging KCs in the course. We assume that these are KCs with the lowest average correctness, $AC(kc_j)$.


%% file: 40-results.tex
\section{Results}
\label{section:results}

\subsection{Impact of Knowledge Components on Assignment Performance}

\subsubsection{Results for Experiment 1.1.: How does the number
of KCs affect the performance on the assignment?} 

\autoref{tab:correctness-vs-kc-num} shows the Pearson correlation between the number of KCs and the assignment performance (average correctness of assignment submissions) $AC$ for the tested datasets. We can see that there is a negative correlation between the number of KCs and the average student performance per assignment for all tested datasets, and for all extracted KC types. This holds for KCs given by an instructor, as well as ones generated by an LLM. This is not an unusual finding, as it is intuitive that the more concepts an assignment addresses, the more difficult it is, but we use this as a confirmation to motivate the proposed assignment analysis.

\input{42-table-correctness-vs-kc-num}

\begin{figure}
    \centering
    \includegraphics[width=1.0\linewidth]{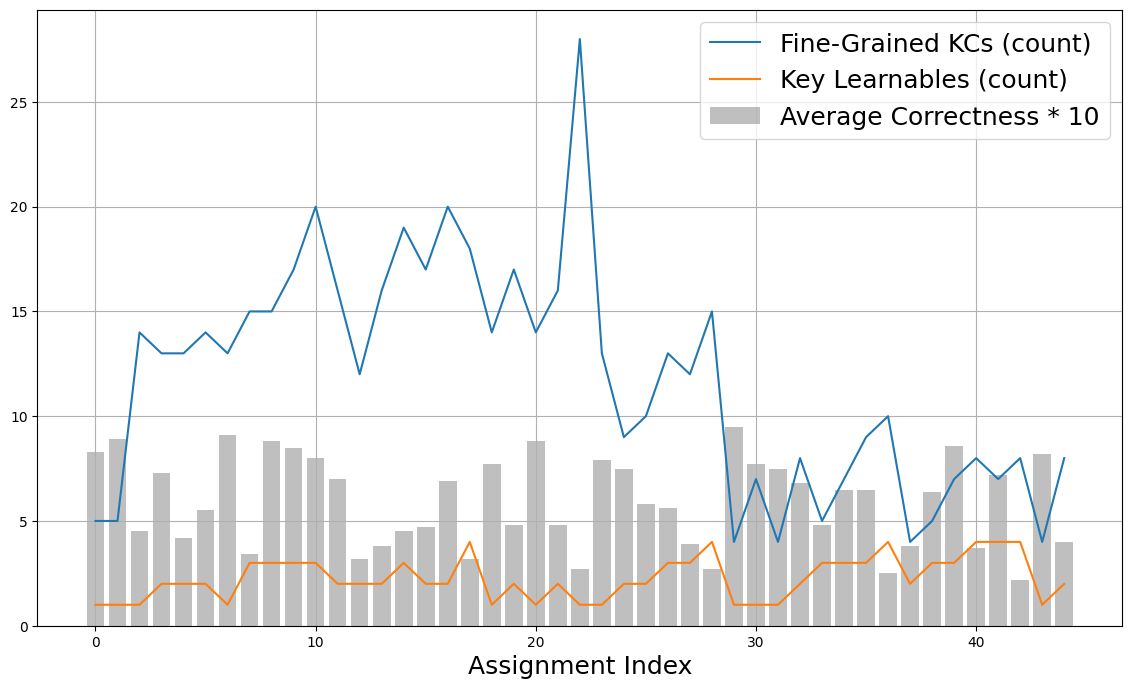}
    \caption{Number of KCs vs assignment performance on the Dart-Intro dataset. The average correctness is multiplied by 10 for the visualization.}
    \label{fig:kc-num-vs-correctness}
\end{figure}

\subsubsection{Results for Experiment 1.2.: How does change of KCs changes the assignment performance?}


In this experiment, we compare the change in KCs - i.e. the number of added and removed KCs to the change in average correctness $AC_{Diff}$.
The results in \autoref{tab:corrdiff-vs-kc-change} show a statistically significant correlation between the number of removed KCs and the increase in assignment performance in the \textit{Dart-Intro}. For this dataset, we also notice negative correlation (although not statistically significant) between the instructor-given KCs and the change in assignment performance. However, this is not observed for the LLM-generated KCs. One potential reason is the granularity of LLM-generated KCs, which appears to affect both the strength and significance of the KC-performance correlation.
For the \textit{Data-Structures} dataset, there is a negative correlation between the number of added KCs and change in assignment performance, but the number of removed KCs is not correlated with it. The reason for this could lie in the difference between the type of the assignments in the two datasets (we will expand on this in the discussion in Section~\ref{section:discussion}).
For the \textit{ITAP Goal} dataset we only had information about student submissions, and we do not know in which order the assignments were presented in the course, therefore we did not attempt a similar analysis for it.

\input{43-table-corrdiff-vs-kc-diff}

\subsection{Analysis of Assignment Quality based on Knowledge Components}

\input{44-fig-kc-change-vs-correctness}

\subsubsection{Identification of Potentially Problematic Assignments.}

To identify \textit{potentially problematic assignments}, we identify assignments for which the correctness drops from the previous assignments, and no new KCs were added. \autoref{fig:kc_change_vs_correctness} shows comparison between the change in correctness and the addition and removal of KCs. 
We search for assignments for which no new KCs have been introduced ($NKC_{Added} = 0$ for any of the KC type),
but the student performance on the assignment drops (i.e. $AC_{Diff}(a_i) < 0$ ). Such assignments are marked with red bars in the figure. 
The plot shows a comparison of the change in correctness (in the gray bars) and the change in KCs per assignment since the previous assignment. 
In the figure, the average correctness is multiplied by 10 for better visual comparison to the numbers of KC change, since we are interested in the direction and the magnitude of these values.
The plot shows more fluctuation in the performance change between the consecutive course assignments, and a higher number of potentially problematic assignments in the \textit{Dart-Intro} dataset. In contrast, the \textit{Data-Structures} course shows less fluctuations between the assignments. This could be explained by the different nature of the two courses. The multiple choice questions in the \textit{Data-Structures} course could be of a more uniform difficulty, in contrast to the programming assignments in the introductory programming course, where the assignments require writing code. 


\subsubsection{Manual Analysis of Potentially Problematic Assignments}

We then manually analyze the incorrect submissions of the assignments that were flagged as potentially problematic in the previous step. 


From manual analysis of incorrect submissions of these assignments in the \textit{Dart-Intro} course, we identified several common problems, such as:
    \textit{Too strict formatting.} Very often the submissions were otherwise correct, but missing a comma or full stop.
    \textit{Use of incorrect symbols} in the names of the variables in the code (such as ä, ö, etc.).
    \textit{Unclear assignment.} Many errors resulted from misunderstanding the task and not necessarily not understanding the concepts being taught.

\begin{figure}[ht!]
    \centering
    \includegraphics[width=0.8\linewidth, alt={Example assignment figure showing a case with no KC changes and a drop in correctness, with common errors such as repeated names and missing punctuation.}]{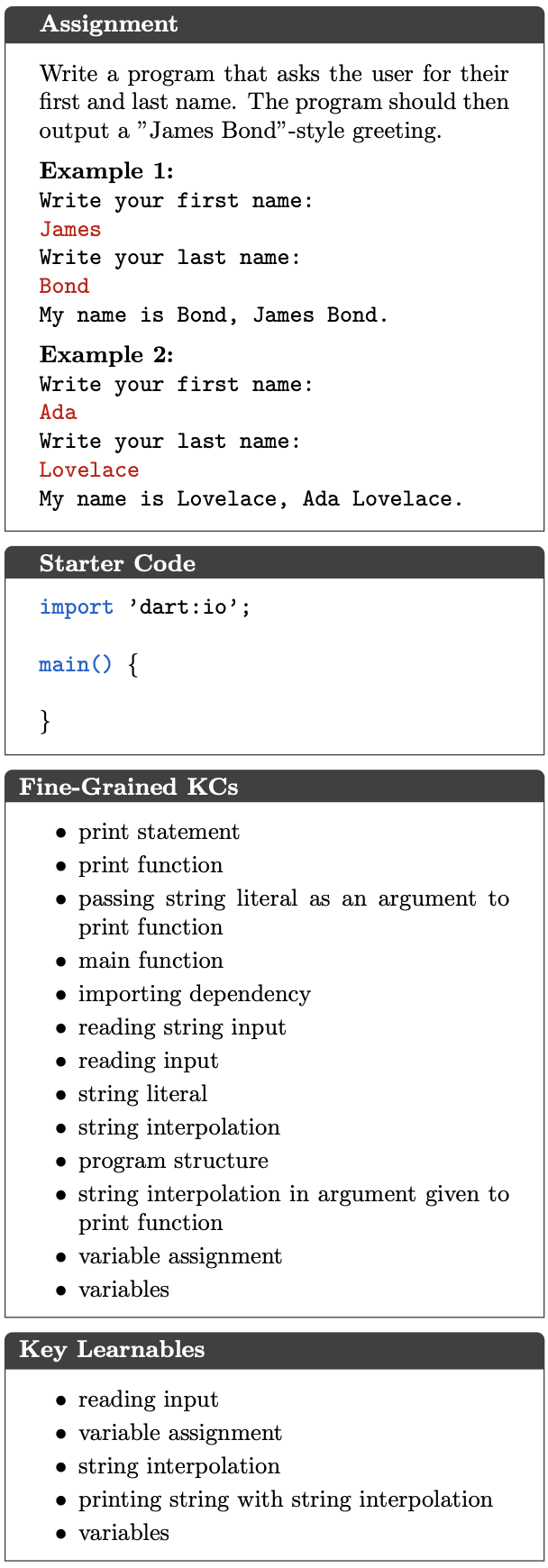}
    \caption{Example assignment, for which no KCs were added or removed since the previous assignment, and the average correctness fell with -0.31, so it was marked as potentially problematic. Manual analysis of the submissions showed that a common error was reading and outputting the name twice, instead of once, as in the correct examples in the figure. Also, many of the incorrect submissions were related to missing comma or full stop.}
    \label{fig:example_assignment}
\end{figure}

\textbf{Example}: The assignment in \autoref{fig:example_assignment} has zero added fine-grained KCs and a difference in the average correctness $-0.31$. After a manual check of the submissions, we identified that many of them read and output the names twice (i.e., reading the name, and printing the pattern \textit{"My name is [lastname], [firstname] [lastname]"} twice).
Many submissions were missing a comma or full stop.




The few assignments marked in red for the \textit{Data-Structures} were examined by the course instructor, but no major problems were found with them.
This suggests that our method can identify potential structural issues in course assignments when they are present, though such issues may not occur in every course.

\subsubsection{Identification of Most Challenging KCs}

For the \textit{Dart-Intro} dataset, we show the KCs with the lowest average correctness of two types--Fine-Grained KCs (\autoref{tab:most-challenging-kcs-low-level-dart}) and Key Learnables (\autoref{tab:most-challenging-key-learnables-dart}), showing average correctness, number of assignments, submissions and users for five KCs with the lowest correctness rate. Interestingly, many of the fine-grained KCs that have the lowest average correctness rate are in line with KCs that one would see in the Rainfall problem, which is known to be challenging for students~\cite{seppala2015we, fisler2014recurring}. 



\begin{table}[ht]
\small
\centering
\caption{Fine-Grained KCs in the Dart-Intro dataset with the lowest average correctness rate. 
Columns are: Average Correctness, Number of Assignments, Submissions and Users.
}
\begin{tabular}{p{4cm}|r|r|r|r}
 \textbf{Fine-Grained KC} & AC & NA & NS & NU \\
 \toprule
summing accumulator in for loop & 0.22 & 2.00 & 9770 & 2141 \\
\midrule
comparing string to parameter value & 0.25 & 2.00 & 9463 & 2368 \\
\midrule
conditional calculation of average & 0.27 & 2.00 & 9383 & 2517 \\
\midrule
incrementing sum in while loop & 0.27 & 2.00 & 9383 & 2517 \\
\midrule
reading input in for loop & 0.27 & 2.00 & 10777 & 2919 \\
\bottomrule
    \end{tabular}
    \label{tab:most-challenging-kcs-low-level-dart}
\end{table}


\begin{table}[ht]
\centering
\small
\caption{Key Learnables in the Dart-Intro dataset with the lowest average correctness rate. 
Columns are: Average Correctness, Number of Assignments, Submissions and Users.
}
\begin{tabular}{l|r|r|r|r}
\textbf{Key Learnable} & AC & NA & NS & NU \\
\toprule
conditional arithmetics & 0.32 & 4.00 & 27589 & 3372 \\
\midrule
nested conditional printing & 0.32 & 2.00 & 10032 & 3235 \\
\midrule
conditional standard input & 0.32 & 2.00 & 10032 & 3235 \\
\midrule
arithmetics in loop & 0.38 & 4.00 & 21934 & 2769 \\
\midrule
standard numeric input in loop & 0.38 & 4.00 & 21934 & 2769 \\
\bottomrule

    \end{tabular}
    \label{tab:most-challenging-key-learnables-dart}
\end{table}



%% file: 42-table-correctness-vs-kc-num.tex
\begin{table}[]
\centering
\caption{Pearson correlation between number of KCs and assignment performance. Statistically significant values (p $<$ 0.05) are marked in bold (p-values in brackets).}
    
\begin{tabular}{@{}lllr@{}}
\toprule
Dataset & KC Type & KC Origin & $AC$ vs $NKC$ \\
\midrule
Dart-Intro      & fine-grained       & expert & \textcolor{neg}{-0.27} (0.072) \\
Dart-Intro      & key learnables  & expert & \textcolor{neg}{\textbf{-0.46}} (\textbf{0.001}) \\
Dart-Intro      & generated       & LLM    & \textcolor{neg}{\textbf{-0.31}} (\textbf{0.036}) \\
Data-Structures & generated       & LLM    & \textcolor{neg}{-0.25} (0.065) \\
ITAP-Goal       & Intro CS        & expert & \textcolor{neg}{\textbf{-0.44}} (\textbf{0.006}) \\
ITAP-Goal       & Tokens          & expert & \textcolor{neg}{\textbf{-0.39}} (\textbf{0.016}) \\
ITAP-Goal       & Tokens Merged   & expert & \textcolor{neg}{\textbf{-0.44}} (\textbf{0.006}) \\
\bottomrule
\end{tabular}

\label{tab:correctness-vs-kc-num}
\end{table}

%% file: 43-table-corrdiff-vs-kc-diff.tex
\begin{table*}[]
\centering
\caption{Pearson correlation between change of KCs (Fine-Graded KCs and Key Learnables): added and removed and change in assignment performance (p-values in brackets). Statistically significant values (p $<$ 0.05) are bolded.}
\begin{tabular}{@{}lllrrr@{}}
\toprule
Dataset & KC Type & KC Origin & $AC_{Diff}$ vs $NKC_{Added}$ & $AC_{Diff}$ vs $NKC_{Removed}$ \\
\midrule
Dart-Intro      & fine-grained      & expert & \textcolor{neg}{-0.24} (0.121) & \textcolor{pos}{\textbf{0.36}} (\textbf{0.017}) \\
Dart-Intro      & key learnables & expert & \textcolor{neg}{-0.29} (0.057) & \textcolor{pos}{\textbf{0.44}} (\textbf{0.003}) \\
Dart-Intro      & generated      & LLM    & \textcolor{pos}{0.02} (0.911)  & \textcolor{pos}{\textbf{0.35}} (\textbf{0.020}) \\
Data-Structures & generated      & LLM    & \textcolor{neg}{-0.17} (0.221) & \textcolor{neg}{-0.08} (0.550) \\
\bottomrule
\end{tabular}

\label{tab:corrdiff-vs-kc-change}
\end{table*}

%% file: 44-fig-kc-change-vs-correctness.tex
\begin{figure*}[]
    \centering


    \begin{minipage}{0.8\linewidth}
    \centering
    \includegraphics[width=\linewidth]{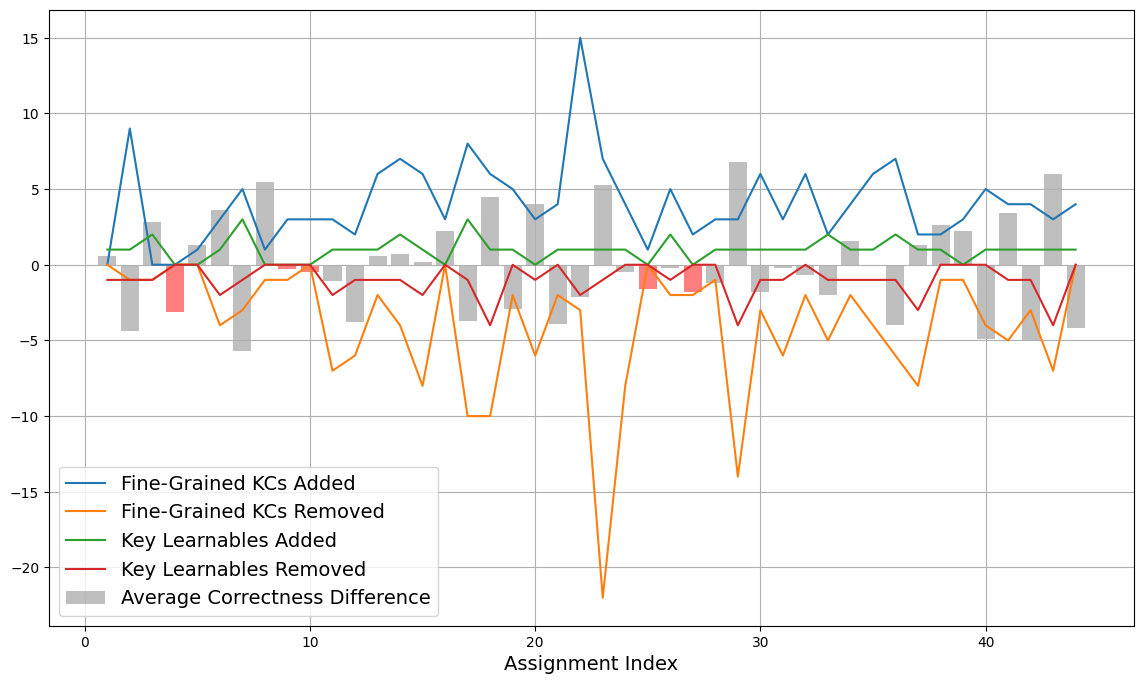}
    \textbf{Dataset \textit{Dart-Intro}, instructor-assigned KCs}
    \end{minipage}%
    \hfill
    \begin{minipage}{0.8\linewidth}
    \centering
    \includegraphics[width=\linewidth]{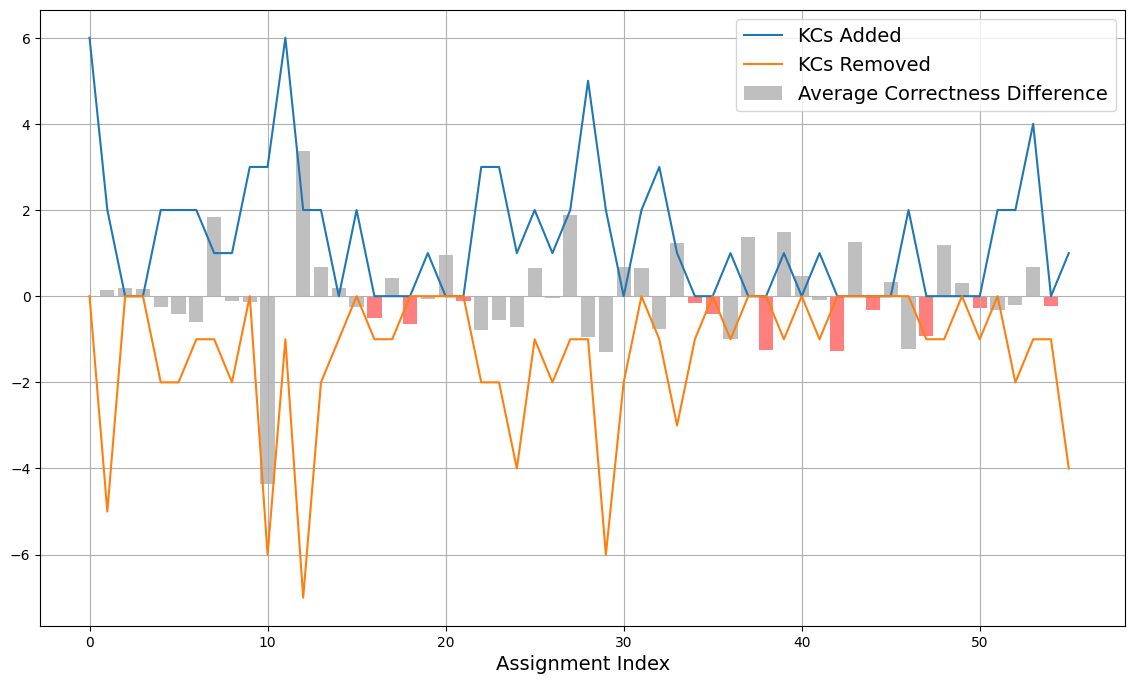}
    \textbf{Dataset \textit{Data-Structures}, LLM-extracted KCs}
    \end{minipage}

    \caption{Identification of \textit{potentially problematic assignments}, marked with red bars.
    The assignment index indicates the order in which the assignments were presented in the course. 
    \textbf{Left}: Assignments are from the Dart-Intro dataset, and the shown KCs are assigned from the course instructor.
    \textbf{Right}: Assignments are from the Data-Structures course, and the KCs are LLM-generated.
    }
    \label{fig:kc_change_vs_correctness}
\end{figure*}

%% file: 50-discussion.tex
\section{Discussion}
\label{section:discussion}


We explore the use of KCs to predict the difficulty of assignments. This perspective complements existing KC-based learner modeling approaches by shifting the unit of analysis from students to assignments.

First, we show that there is a statistically significant negative correlation between the number of knowledge components and the performance on the assignment. This can be intuitively interpreted as \textit{the more topics an assignment is trying to teach, the harder it is}. 

Then we investigate how the change in KCs affects the change in performance in two consecutive assignments. The results indicate that when KCs are removed, the assignment performance improves. Although neither of these results is \textit{surprising}, they highlight that KCs can be a useful method for evaluating the difficulty of programming exercises.

We also propose a framework for using the number of KCs to identify potentially problematic assignments, following the intuition that if new KCs are added, the performance on the assignment drops, and the opposite - if no new KCs were added, i.e., if no new concepts are being taught with an assignment, the performance should increase. Following this intuition, we identify assignments for which no new KCs (of at least one of the levels) were added, but there is a negative change in performance. We then show that a manual examination of these assignments can help identify common errors with the assignment itself, which are not necessarily related to the concepts being taught. Such problems can be related to formatting requirements that are too strict, the use of forbidden symbols in the code, or unclear requirements.

By comparing two datasets, one consisting of introductory programming assignments and another of multiple-choice questions on data structures, we show that our method can reveal potential issues, though these may vary across courses or depend on the nature of the assignments.

\section{Limitations and Threats to Validity}

In the analysis of how the change in KCs affects the change in assignment performance, we rely on the fact that assignments are ordered. However, in some courses, it is possible that some students can start from different exercises or go back and forth.
To extend this analysis to a course format where the assignments are not necessarily presented in a fixed order but the students can choose the order of assignments, adjustments need to be made, such as modeling assignment sequences at the student level, accounting for varying learning paths, and redefining KC transitions based on individual progression rather than a fixed global order. 

The correlation between the absolute number of KCs and assignment performance (RQ1) is verified on three datasets, and is shown to be statistically significant on two of them.
The analysis on KC addition and removal and its correlation with performance (RQ2) is based on two datasets, and is not always consistent across the datasets and KC types. While we consider the findings in this study insightful for determining the relations between assignment complexity and number of KCs, it would be interesting to see how they generalize to even more courses, disciplines, programming languages or assignment types.


Each assignment typically had more than one KC. We only considered the raw numbers of KCs per assignment and their relationship with assignment complexity, but the interactions between the KCs could be more complex than just the number of them.

The proposed analysis is affected by the quality of the generated KCs. This could depend on how carefully the instructor selected the KCs, what they took into account when doing the KCs tagging, and the consistency of the granularity throughout all the assignments. When the KCs are extracted with and LLM, the quality can depend on how well the LLM performed and the prompt type. 
However, the topic of KC quality is out of scope of this work, and is being addressed in separate line of research. Our approach assumes good-quality KCs, which could be used for other purposes, such as knowledge tracing. We rely on the fact that KCs could already be available, and our method could utilise them as an additional analysis of the quality of the materials. 


%% file: 60-conclusion.tex
\section{Conclusions and Future Work}
\label{section:conclusion}

In this paper, we explore how the performance of programming assignments relates to the number of knowledge components in an introductory programming course and a course on data structures. 
We show that there was a correlation between the number of KCs and the performance on the assignment, and this is valid for different levels of KCs, and also for different origin of the KCs--given by a course designer or extracted with an LLM.
These findings can serve as a suggestion for course creators to design their courses such that the introduction of new KCs is balanced and gradual--to not change drastically the number of new KCs from one assignment to the other, i.e., to not introduce too many new concepts at once.
This perspective is also aligned with prior work on personalized programming practice and recommendation, where KC progression is used to guide learners through assignments and support informed navigation decisions \cite{RN5686, RN5964}. In such systems, KC-based signals can be visualized to help students choose appropriate next steps.

We propose a method to identify potentially problematic assignments, such as those where no new KCs were added but performance declined. Manual inspection of such cases can reveal issues unrelated to conceptual understanding, like unclear instructions or overly strict output formats. 

Our method can use KCs defined by the instructor or extracted automatically by an LLM, offering instructors an additional way to assess assignment quality beyond average correctness. It can be applied to any course with ordered assignments and measurable performance, 
though not all courses may contain issues unexplained by the number of introduced concepts.





A possible direction for future research is to perform a lagged analysis of how previous KC load influences student dropout, examining whether consecutive cognitively demanding exercises contribute to higher attrition. The proposed method could also be adapted to handle assignments presented in a non-sequential order and to utilize the relationships among KCs.